\documentstyle[aps,prb,epsf,rotate]{revtex}

\makeatletter
\let\chapter\hid@chapter
\makeatother
\newcommand{\ph}{\hat {\bf p}}
\newcommand{\nh}{\hat {\bf n}}
\newcommand{\vvf}{{\bf v}_f}
\newcommand{\gh}{\hat g}
\newcommand{\Dh}{\hat \Delta}
\newcommand{\Rvec}{{\bf R}}
\newcommand{\that}{\hat \tau}
\newcommand{\mat}{\epsilon_n}

\newcommand{\intpp}{\int{{d\Omega_{\ph^\prime}}\over {4\pi}}\;}
\newcommand{\intphi}{\int_> {{d\phi}\over {2\pi}}}
\newcommand{\msum}{\sum_{\mat}}
\newcommand{\be}{\begin{equation}}
\newcommand{\ee}{\end{equation}}
\newcommand{\bdm}{\begin{displaymath}}
\newcommand{\edm}{\end{displaymath}}
\newcommand{\ba}{\begin{eqnarray}}
\newcommand{\ea}{\end{eqnarray}}
\begin{document}
\begin{title}
{ \bf Pinhole junctions in d-wave superconductors}
\end{title}
\author{Mikael Fogelstr\"om$^{a,b}$, Sungkit Yip$^{a}$
       \footnote{corresponding author, fax: +1 847-491-9982, 
        e-mail: yip@snowmass.phys.nwu.edu}
        and Juhani Kurkij\"arvi$^{a,b}$}
\address{ $^{a}$Department of Physics \& Astronomy, 
           Northwestern University, Evanston, Illinois 60208, U. S. A. \\
  $^{b}$Department of Physics, {\AA}bo Akademi, 
           Porthansgatan 3, 20500 {\AA}bo, Finland} 
\date{Accepted for publication in Physica C}
\maketitle
%
%
\begin{abstract}
We present a self consistent treatment of pinhole junctions in 
$d_{x^2-y^2}$ superconductors. 
The current-phase relation $j_s(\chi)$ is 
studied at different temperatures and at different angles 
$\alpha$ between the crystal $\hat a$-axis and the junction 
(surface) normal. We show that the 
critical current of a junction can be reduced 
by pair-breaking effects at the separation.  We also study the 
Josephson energy 
of a pinhole as a function of the phase difference across the 
junction. 
In particular are mapped the positions of the energy 
minima $\chi_{min}$ 
at different temperatures as functions 
of ($\alpha_L,\alpha_R$), the crystal orientations of the left 
and right superconductors. 
With decreasing temperature
there is an increasing range of crystal orientations where 
$\chi_{min}$ varies continuously from 0 to $\pi$. 


keywords: superconducting junctions, unconventional pairing 
\end{abstract}
%
%
\section{Introduction}
\label{sec:intro}

Recent experiments and their interpretations have strongly argued 
in favor of an order parameter of $B_{1g}$ ( $d_{x^2-y^2}$ )
symmetry in the oxide
superconductors (see e.g. reviews \cite{Scalapino,Annett96}). Part
of the evidence comes from bulk 
and transport properties (see \cite{Scalapino}). Other
relevant data involve Josephson junctions which
probe the order parameter at surfaces. In these latter experiments,
the phase of the order parameter is involved, and "interference"
experiments can be carried out. 

It will be  important below to realize that interference experiments fall
into two categories. 
In the first one measures the critical current of a two junction
loop as a function of the magnetic flux treading the ring and records
the positions of the current maxima
\cite{Wollman93,Brawner94,Iguchi94,Wollman95,Miller95,Chaudhari94}. The second
type consists of mapping the phase differences at the Josephson
energy minima
\cite{Tsuei94,Kirtley95,Mathai95,Tsuei96,Kirtley96}.

Josephson junctions in s-wave superconductors have been well explored.
Here we distinguish  between "tunnel" junctions on the
one hand and "weak-links" on the other. We shall follow
\cite{Likharev79} and classify junctions on the basis of their {\it
normal-state} conductivities. In a tunnel junction normal electrons
face a reasonably high potential barrier whereas in a weak-link
junction they do not. 

Even in s-wave superconductors, tunnel junctions and 
weak links behave
in different fashions. We compare  weak-link junctions with tunnel
junctions in unconventional superconductors 
picking the pinhole as the
representative of the former.  
The differences arise from two sources:
Firstly, the current-phase relation of a tunnel junction is
sinusoidal. The minima/maxima  of the Josephson energy therefore lie
at a phase difference equal to zero or $\pi$ (provided the order
parameter does not break time-reversal  symmetry \cite{Yip95}). In a
weak link, the current-phase relationship  is not sinusoidal 
in general \cite{Yip95,Barash95}, and the 
minima/maxima will not lie at zero or
$\pi$. Secondly, in weak links the current carrying processes
sample a wider angular range of momentum than in tunnel
junctions. With  different crystal orientations on the two sides of a
junction, varying  the angle of the junction with respect to the
crystals, one observes quite different behaviors of the positions of
the energy minima and critical current in pinholes and in tunnel
junctions.  

Nevertheless, "weak-links", (especially our pinhole) and tunnel
junctions are the opposite ends of a continuum. Imagining 
a weak potential barrier in the link and increasing its strength
should take one smoothly from the 
pinhole to the tunnel junction. 

Josephson junctions are  probes of surfaces.  
Near a surface, an unconventional order parameter
must be strongly affected \cite{AGR,Buchholtz95}. 
We therefore also investigate the effect of surface pair breaking
on our weak links and tunnel junctions.

Most of the discussions on high-Tc junctions in the literature
ignore the difference between weak links and
tunnel junctions (and the effects of surface pair-breaking). 
In the sequel it is argued in some detail that a number of
puzzling reports in the literature may
have a natural explanation if one keeps track of these
effects.  One should be very careful when drawing
conclusions on the order parameter type of
a high $T_c$ superconductor on
the basis of measurements of various Josephson-related
effects.  All available models and their variants 
should checked out for competing interpretations.

In Section 2 we discuss the self-consistently computed
junctions, first pinholes (weak links) and then tunnel
junctions in Section 3. Section 4 is about the non
self-consistent junctions, i.e. those treated analytically
assuming constant order parameters up to the separation
where the junction sits.  Section 4 ends with the Subsection
4.1 which compares the self-consistent and non
self-consistent models introduced in the previous Sections. 
Section 5 is on critical currents and phase differences where
the Josephson energy reaches its minima.  The final
Section 6 is on the conclusions we suggest from experiments. 
The ultimate Section 6.3 is a critical analysis of
the plausibility of our models as describing real
junctions.
 \footnote{We shall only consider planar interfaces.
Some recent papers ( e.g. \cite{Mannhart96} and references therein)
have suggested that it may be important to include the effects of facets
in understanding the properties of some grain boundaries. 
 If this is so, then we cannot directly apply our results
to these grain boundaries; instead they can only serve   
as inputs when considering the facets. }

%
\section{The pinhole junction}
\label{sec:calc}
\begin{figure}
\centerline{
\epsfysize=.5\textwidth 
\rotate[r]{
\epsfbox{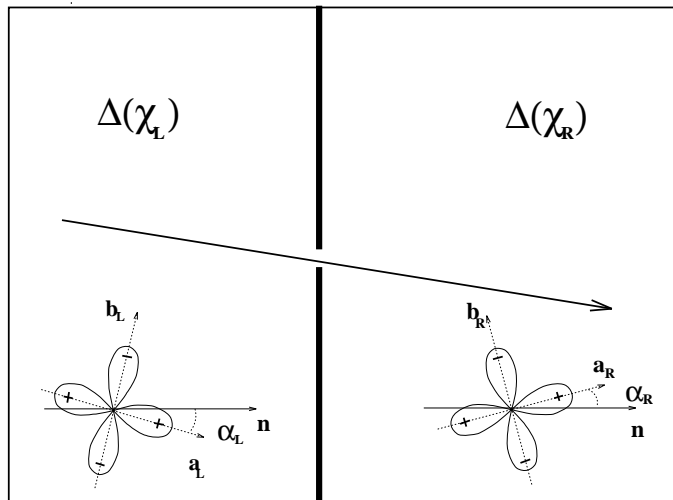} }}
\caption[]{Sketch of a pinhole. The opening is small on the
scale of 
coherence length. Most trajectories are
reflected at the interface, and only those hitting the hole
are transmitted. } 
\label{fig:pinhole}
\end{figure}

The pinhole junction is a small opening in an interface 
separating two superconductors (see Figure \ref{fig:pinhole}), small
both in length and  in width on the scale of the coherence 
length $\xi_0$. Being so tiny, the opening perturbs little
the order parameter which may 
be calculated ignoring the pinhole. A phase difference $\chi$ 
over the junction is introduced 
multiplying the self-consistent order parameters on the
two sides with the factors 
$\exp(\pm i\chi/2)$. The phase is taken to hop
discontinuously at the partition. The pinhole junction 
was first introduced in s-wave 
superconductors \cite{Kulik77} and has also been considered as a
model of a weak link in superfluid $^3$He \cite{Kopnin86,Kurkijarvi88}.

\subsection{The self-consistent pinhole}
\label{subsec:selfpinhole}
Surfaces have a pair-breaking effect on a
superconductor whose
order parameter depends on the direction $\ph$ of propagation
of a quasiparticle, and we need to take into account
the quasiparticle scattering at the wall. Models for 
walls have been studied in $^3$He (see, for instance,
\cite{Kurkijarvi87} and \cite{Fogelstrom96}) and recently also in
d-wave superconductors (\cite{Buchholtz95}, \cite{Matsumoto95}
and  \cite{Zhang95}). We chose the specular surface, 
attractive in its simplicity.
It has been studied 
by \cite{Buchholtz95}. The Green's function 
$\gh(\ph,\Rvec;\mat)$ in the specular model
is taken to be continuous at the surface along
pairs of trajectories $(\ph_{in},\ph_{out})$  connected as 
\begin{equation} 
\label{trajpair}
\ph_{out}=\ph_{in}-2\nh (\nh\cdot\ph_{in}),
\end{equation}
$\nh$ being the surface normal.

With the boundary conditions set, the procedure of getting at the
order parameter consists of iterating the Eilenberger equation 
\begin{equation}
\label{Eilenberger}
\lbrack i\mat\that_3-\Dh(\ph,\Rvec),\gh(\ph,\Rvec;\mat)\rbrack
+i\vvf\cdot\nabla_{\Rvec}\gh(\ph,\Rvec;\mat)=0
\end{equation}
for the propagator,
where $\vvf$ is the Fermi velocity and $\ph$ the direction
of momentum at the Fermi surface,
together with the gap equation for the order parameter 
\begin{equation}
\label{gapequation}
\Delta(\ph,\Rvec)=\pi T \msum\intpp V(\ph,\ph^\prime){1\over2}
{\rm Tr}\lbrace (\that_1-i\that_2) \gh(\ph^\prime,\Rvec;\mat)\rbrace
\end{equation}
till self consistency. As indicated by the carets, 
the propagator $\gh$ and the order
parameter $\Dh$ are matrices in the particle-hole space 
\footnote{\label{foot:decomp}
The propagator is decomposed as
$\gh=\sum_{i=1}^3 g_i \that_i$,$\that_i$ being the Pauli
matrices .}.
The order parameter is conveniently parameterized 
in terms of a real part $\Delta_2$ and an 
imaginary part $\Delta_1$ as
$\Dh=i(\Delta_1\,\that_1+\Delta_2\,\that_2)$. Only interested in 
an equilibrium property, we can use the Matsubara formalism at the
Matsubara frequencies $\mat=\pi T(2n+1)$. 

The pairing interaction $V(\ph,\ph^\prime)$ in equation 
(\ref{gapequation}) determines the symmetry 
of the order parameter. $V(\ph,\ph^\prime)$ can be separated into a
sum of allowed pairing channels. The strength of a
particular channel $X$ depends on the coupling parameter 
${1\over{V(X)}}$ which
can be eliminated in favor of the transition temperature
$T_c(X)$ in the channel with the aid of the well-known BCS-relation
\begin{equation}
\label{intstrength}
{1\over{V(X)}}=\ln {T\over{T_c(X)}}+\sum_{n\ge0}^{n_c}
{1\over{n+{1\over2}}}
\end{equation}
where ${n_c}$ is a cutoff. In this work, the dominant pairing
channel is always chosen as having an order parameter of $B_{1g}$
symmetry. Buchholtz et al \cite{Buchholtz95} have shown that for
equal $V(X)$ in the subdominant channels, $B_{2g}$ has the strongest
effect on the properties of a smooth surface. Thus
we follow \cite{Buchholtz95} and study a superconductor with a
$d_{x^2-y^2}$ or $B_{1g}$ order parameter in the bulk and allow an
admixture of a $d_{xy}$ or $B_{2g}$ component close to the
surface. The degree of mixing is controlled by the ratio of the
transition temperatures of the two representations. This ratio is 
a parameter of the present calculation. The pairing interaction is
\begin{equation}
\label{pairing}
V(\ph,\ph^\prime)=2 V_{B_{1g}} \cos 2\phi \cos 2\phi^\prime
                 +2 V_{B_{2g}} \sin 2\phi \sin 2\phi^\prime.
\end{equation}

We refer to \cite{Buchholtz95} for a full account of 
the behavior of the order parameter at a surface and only 
briefly summarize the principal 
effects here. 
Imagine a superconducting crystal with a $d_{x^2-y^2}$ 
order parameter cut with a surface whose normal is $\nh$. In
a thin film sample, $\nh$ would lie normal to the junction 
in the plane of
the film. The d-wave order parameter is reduced when the crystal 
is rotated relative the surface (junction) normal $\nh$ to a 
position 
determined by the angle $\alpha$ between the crystal $\hat a$-axis 
and the normal $\nh$ (see Fig. \ref{fig:pinhole})
\footnote{We take the d$_{x^2-y^2}$ order parameter as having its 
positive lobe along the the crystal $\hat a$ axis 
and its negative lobe
along the $\hat b$-axis.}. 
The order parameter is unaffected by the presence of the
surface (=cut) if the crystal has its $\hat a$ or 
$\hat b$-axis parallel to $\nh$. It is maximally  
reduced when $\alpha$ is equal to ${{\pi \over 4} +n{\pi \over 2}}$, 
n being an integer. 
In the latter case, the order parameter vanishes 
identically at the wall. 
If the ratio of transition temperatures is 
chosen larger than zero, a $B_{2g}$ 
component may develop in the vicinity of the wall. 
A special case is at $\alpha$ close to ${\pi\over 4}$ 
at temperatures $T/T_c(B_{1g})\le T_c(B_{2g})/T_c(B_{1g})$ where
the composite order parameter is found to break 
time-reversal symmetry \cite{Matsumoto95,Fogelstrom97}.

There is, however, a problem associated with mixing two 
representations as described above. On the basis of 
the boundary condition 
$\gh(\ph_{in},{\bf 0};\mat)=\gh(\ph_{out},{\bf0};\mat)$, 
eq (\ref{gapequation})
and the general symmetries of the propagator\footnote{For symmetries 
obeyed by propagators and self energies, see
\cite{Serene83}.} one can show that (see also \cite{Palumbo96}), 
for $T_c(B_{2g})$ different from zero, the ratio of the two 
order parameter components at the wall is\footnote{It seems
 that eq. \ref{opratio1} must reflect itself in  the boundary condition
that should be used in a phenomenological Ginzburg-Landau theory.
It is unclear whether this has been 
taken into account properly in the literature 
(e.g. \cite{Sigrist95})}
\begin{equation}
\label{opratio1}
{{\Delta_{B_{2g}}(0)}\over{\Delta_{B_{1g}}(0)}}
=-  { V_{B_{2g}} \over V_{B_{1g}} } \tan 2 \alpha. 
\end{equation}
Numerically this becomes
\begin{equation}
\label{opratio}
{{\Delta_{B_{2g}}(0)}\over{\Delta_{B_{1g}}(0)}}
=-\tan 2 \alpha \Biggl(1+{{\ln {{T_c(B_{1g})}\over{T_c(B_{2g})}}}\over
{\ln {T\over{T_c(B_{1g})}}+
\sum_{n\ge0}^{n_c}\;{1\over {n+{1\over 2} }}}}\Biggr)
\end{equation}
i.e. the ratio is cutoff dependent. We must 
keep track of the the cutoff as
soon as the $T_c$-ratio is non-zero. The results given below are all
at the cutoff $n_c (T/T_c)$ chosen equal to 
the integer part of $16 T_c/T$.

With the order parameter determined self-consistently, 
we can calculate the current-phase relation across a junction. 
For this we need the Green's function at the orifice. The propagator 
is computed along trajectories through 
the orifice (see Figure \ref{fig:pinhole}). The boundary condition
is boundedness far away on
both sides of the interface.  At the orifice, the propagator is
matched for continuity. An alternative route 
is the multiplication trick \cite{Thuneberg84}.
This slight of hand takes advantage of exploding solutions  
along trajectories towards the junction. 
The matrix commutator of each
pair of diverging solutions on the same trajectory at
the pinhole (in fact an
exploding solution and a decaying solution if viewed as
propagating in the same direction) 
delivers the physical propagator in the 
orifice. 
The current density for a given
phase difference $\chi$ is calculated from the propagator in the
orifice  (at $\Rvec=0$) by  
\begin{equation}
\label{gencurrent}
j_s(\chi,T)=2 e N_f\, v_f\,  T\msum\intphi\; \cos \phi\,{\rm Tr_2}
\lbrack \that_3 \gh_{\chi}(\ph,0;\mat)\rbrack.
\end{equation}
Here $N_f$ is the density of states at the Fermi surface, 
and $>$ on
the integral means that only the half sphere $\nh\cdot\ph > 0$ of
directions are included.
In general the Fermi velocity and the density of states depend on
the position on the Fermi surface $\ph$. We assume
a circular Fermi surface, for which the Fermi velocity is
$\vvf=v_f \ph$. This assumption is not an unreasonable simplification.
Buchholtz {\it et~al.}\cite{Buchholtz95} compared calculations
of the order parameter and the surface density of states
using a cylindrical Fermi surface with the same calculations 
using a Fermi surface calculated from a tight-binding model\cite{Radtke92}.
They found very small differences between the two models. 

The phase dependent part of the Josephson energy density 
can be calculated from the current-phase relation as
\begin{equation}
\label{genenergy}
E_s(\chi,T)=E_s(\chi_0,T)+{\hbar \over{2e}}\int^{\chi}_{\chi_0}
d\chi^\prime j_s(\chi^\prime,T).
\end{equation}
Equation (\ref{genenergy}) is valid for all the junctions
discussed in the present article.


\section{The tunnel junction}
\label{sec:tunju}
As announced in the introduction, we wish to compare 
weak link junctions with tunnel junctions. A tunneling barrier is
characterized by its transparency $D(\phi)$. Self consistent or not,
the transparency of the junction may depend on the direction of the
incident momentum ($\phi$ being the angle $\ph$ makes with the
surface normal). We have chosen a transmission coefficient
proportional to $\exp(-8 \sin^2\phi)$, i.e. peaked in the forward
direction. This model transmission is used though out this paper with
one exception. In the initial discussion in Section
\ref{subsec:posenergy} the transmission coefficient is  chosen as
all peaked along the junction normal.  This we refer to as the
forward (backward) tunneling limit. 

\subsection{The self-consistent tunnel junction}
\label{subsec:sctj}
The self-consistent order-parameter calculation
of Section \ref{subsec:selfpinhole} works here
as well since the tunneling potential barrier is
high enough to perturb the order parameters on
the two sides only little. For calculating  the tunneling current we 
take the expression of \cite{Millis88} 
\be 
\label{tuncurrent}
j_s(\chi,T)= {2 e N_f\, v_f\, \over \pi} T\msum \intphi\; D(\phi)\, \cos\phi \,
\bigl\lbrack g^L_2(-\chi) g_1^R(\chi)-g_2^R(\chi) g_1^L(-\chi)\bigr\rbrack 
\ee
where $g_{1,2}^{R,L}$ are the anomalous part of the propagator at the
tunneling barrier (in our case at the surface) in either superconductor. 

%
\begin{figure}
\centerline{
\epsfxsize=0.3\textwidth
\rotate[l]{
\epsfbox{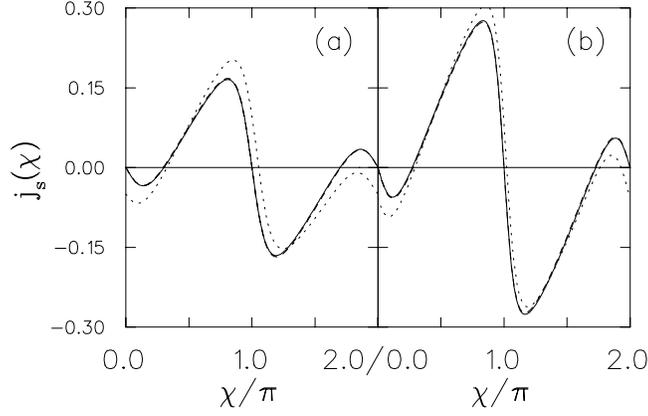}}}
\caption[]{Current-phase relation of a weak link
with $\alpha_L=9 \pi/40$
and $\alpha_R=\pi/40$ and temperature  $0.1\; T_c(B_{1g})$. 
(a) self-consistent current relations for three different
$T_c$-ratios: $0$ (solid line), $10^{-3}$ (dashed line) 
and $10^{-1}$ 
(dotted line).  (b) mimicking the  self-consistent current-phase
relation  via a  mixture of 
$B_{1g}$ and $B_{2g}$. The choices are: 
Pure $B_{1g}$ order 
parameters in both
superconductors  (solid line),  
$\Delta^{R,L}(\phi)=\Delta_{B_{1g}}-0.05\Delta_{B_{2g}}(\phi)$ 
(dashed line) and  $\Delta^L(\phi)=\Delta_{B_{1g}}(\phi)$ and
$\Delta^R(\phi)=\Delta_{B_{1g}}(\phi)-0.05 i\Delta_{B_{2g}}(\phi)$ 
(dotted line).  
(a) and (b) the dotted lines: states breaking
time-reversal symmetry.}
\label{fig:current} 
\end{figure}
%
%
\begin{figure}
\centerline{
\epsfxsize=0.3\textwidth
\rotate[l]{
\epsfbox{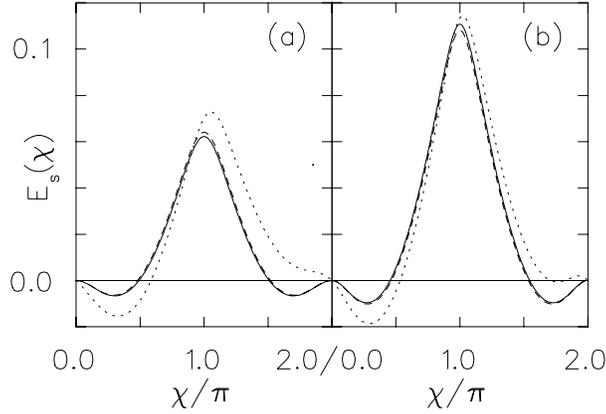}}}
\caption[]{Energy-phase relation for the same 
junction as in figure
\ref{fig:current}. 
Lines as in Fig. \ref{fig:current}}
\label{fig:energy}
\end{figure}
%
%
\section{Non self-consistent junctions}
\label{sec:nscj}
Most of the time we investigate the junctions with 
the order parameter self-consistently calculated.
As a point of comparison, we consider the case
of a constant order parameter up to the surface 
\cite{Yip93,Yip95,Barash95}. This non self-consistent
calculation can usually be done analytically. Comparing with the self
consistent calculations we get a measure on the 
effects of surface pair
breaking on the current-phase and the energy-phase relations. 

Consider a junction, weak link or tunnel junction, with order
parameters left unaffected by scattering at the wall or the 
junction surface. Assuming tetragonal crystal symmetry, the bulk-gap 
function in d-wave superconductors is 
\begin{equation}
\label{dx2y2gap}
\Delta_{B_{1g}}(\ph)=\Delta \cos 2(\phi-\alpha)
\end{equation}
for order parameters of $d_{x^2-y^2}$ or $B_{1g}$ symmetry and 
\begin{equation}
\label{dxygap}
\Delta_{B_{2g}}(\ph)=\Delta \sin 2(\phi-\alpha)
\end{equation}
for order parameter of $d_{xy}$ or $B_{2g}$ symmetry.
$\Delta$ is the temperature dependent maximal amplitude of the 
gap and $\phi$ the angle $\ph$ makes with respect to the
junction normal $\nh$. 

The expressions for the current through the junctions are calculated 
inserting the bulk propagators into equations (\ref{gencurrent}) 
and (\ref{tuncurrent}). The weak link current is given by
\cite{Yip93}, and the tunnel-junction current is 
basically the well known
Ambegaokar-Baratoff formula \cite{Ambegaokar63}. 
\subsection{Self-consistent pinhole versus non self-consistent
pinhole}
\label{subsec:comparing}
We calculate and compare the current-phase 
(Fig. \ref{fig:current}) and the  
energy-phase relations (Fig. \ref{fig:energy}) for a self
consistent (a) and a non-self-consistent (b) weak link at
different $T_c$-ratios at $T= 0.1 T_c$.
The crystal orientations are chosen as 
$\alpha_L=9\pi/40$ and $\alpha_R=\pi/40$. 
We see that even with a pure $B_{1g}$ order parameter, 
the very similar current-phase relationships
differ significantly from the sinusoidal form. The 
energy minima occur at some $\chi$, neither equal to $0$ nor $\pi$.
If we admit, in a self-consistent fashion, 
a small $B_{2g}$-order parameter, 
the critical current increases a little but the energy minima are
not shifted in position. When, on the other hand, 
the $T_c$-ratio is set 
equal to the temperature (dotted line in (a) and (b)), the system is 
just on the edge of the area in the parameter space where
the superconducting state will spontaneously break
time-reversal symmetry near the surface. There
the current-phase relation no longer 
obeys the symmetry $j_s(\chi)=-j_s(-\chi)$ and 
the critical current is
increased. 

Turning to the equally very similar energy-phase relations, 
we see that the symmetry  of the
time-reversal invariant relation has been lost at the edge of the
instability against a time-reversal symmetry breaking state (TRSB).  
This applies also to the mimicked state in the non 
self-consistent case
(dotted line in (b). The junction can in fact be in two distinct
states (with different order parameters).  The second
state has $E_s (\chi) \to E_s ( 2 \pi - \chi)$ 
(not shown).   

An overall conclusion is that the principal effect of pair
braking on a surface reflects itself as a simple reduction
of the critical current.
\begin{figure}
\centerline{
\epsfxsize=0.7\textwidth
\epsfbox{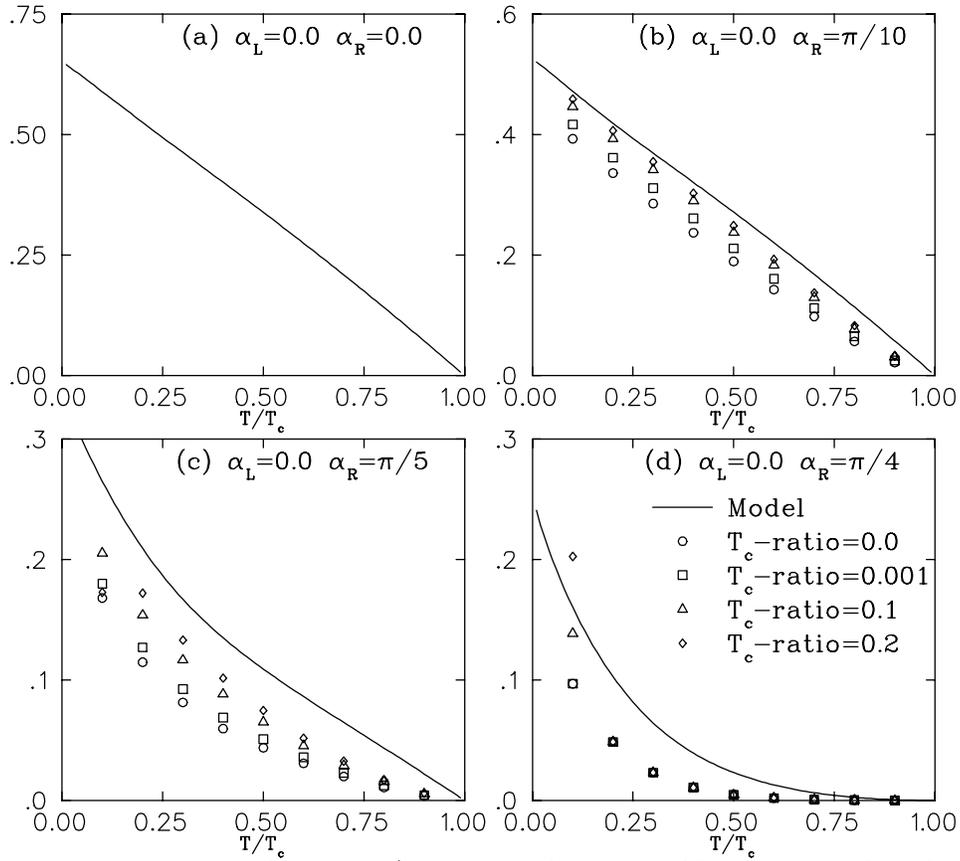} }
\caption[]{Critical current vs. temperature in a
d-wave/d-wave weak link at four different crystal
orientations. Unit of current,  $ 2 \pi T_c / R$.}
\label{fig:wltempcurr}
\end{figure}
\begin{figure}
\centerline{
\epsfxsize=0.7\textwidth
\epsfbox{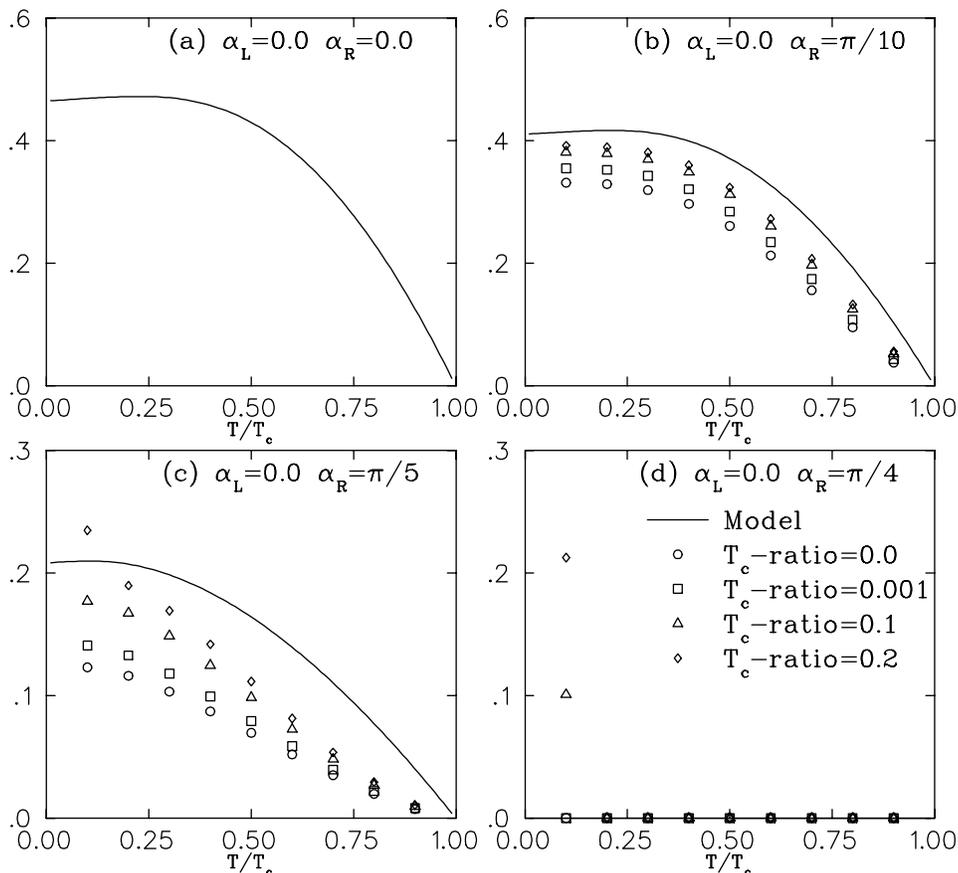} }
\caption[]{Critical current vs. temperature in a 
d-wave/d-wave tunnel junction at four different crystal
orientations. Unit of current, $ 2 \pi T_c / R$.} 
\label{fig:tutempcurr}
\end{figure}

\section{Critical currents and positions of energy minima}
\label{sec:critical}
\subsection{Critical currents}
\label{subsec:critcurrents}
\begin{figure}
\centerline{
\epsfxsize=0.6\textwidth
\rotate[l]{\epsfbox{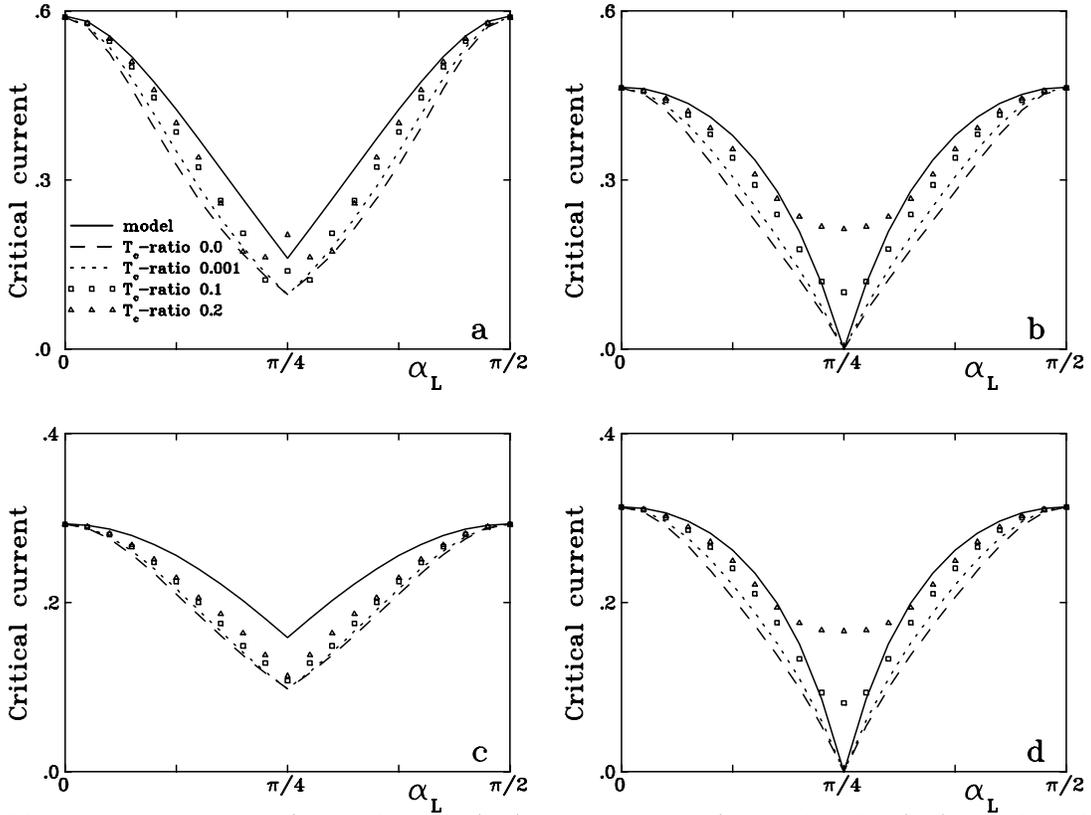}}}
\caption[]{Critical currents of a d-wave/d-wave
junction (a-b) and of an s-wave/d-wave junction
(c-d) vs.
orientation angle $\alpha_L$ at $T=0.1T_c$.  
Weak links, (a),(c); tunnel junctions, (b), (d) .}
\label{fig:anglecurr}
\end{figure}

In Figure \ref{fig:wltempcurr}, the 
critical current of a weak link between two d-wave superconductors at
different crystal misorientations is shown as a function of
the temperature. With the misorientation increasing 
towards $\pi \over 4$ the temperature dependence of the
critical current evolves from linear in $(1-T/T_c)$
at $\alpha_R=0$ to quadratic, $(1-T/T_c)^2$ at
$\alpha_R={\pi\over4}$ for the non-self-consistent (model)
calculation and an even higher power for the self-consistent
one. The solid
line present in Figs.\ref{fig:wltempcurr}-\ref{fig:anglecurr}
indicates again (see Section \ref{subsec:comparing}) that the principal
effect of the surface scattering is reducing the critical current.
Adding a portion of a $B_{2g}$
component compensates for this reduction depending on
the relative strengths of the two representations.
At misorientations close to ${\pi\over 4}$, entering a  TRSB state at
low temperatures may boost the critical current. Below $T_c^{TRSB}$ the
temperature dependence of the critical current undergoes an abrupt
deviation from the $(1-T/T_c)^2$ behavior as clearly seen in Fig. 
\ref{fig:wltempcurr}(d). In other words, the temperature
dependence of the critical current can serve as a detector of a 
TRSB-state at low temperatures. 

In Figure \ref{fig:tutempcurr} the same temperature
dependence as for the weak link in Fig.
\ref{fig:wltempcurr} is shown for a tunnel junction.
Same as with the weak link, we see that the critical
current is reduced by the surface pair breaking. 
Approaching $\alpha_R = {\pi\over 4}$ extends the initial linear 
temperature 
dependence close to $T_c$ to lower temperatures. 
Right at $\alpha_R ={\pi\over 4}$,
the current vanishes as long as the superconducting 
state is not a TRSB-state.

The critical current at $T=0.1 T_c$ as a function of crystal
orientation $\alpha_L$ is shown in Fig. \ref{fig:anglecurr}. 
Here $\alpha_R =0$, i.e, the $\hat a$-axis of one of the
superconductors is parallel to the junction normal.
In Fig.\ref{fig:anglecurr} a-b the junction is between two d-wave
superconductors and  in \ref{fig:anglecurr} c-d between
a d-wave and an s-wave superconductor. Once more the 
TRSB-state 
has a most spectacular effect on the critical current. For the
tunneling junction, the critical current no longer vanishes 
at $\alpha_L={\pi \over 4}$ as in the time-reversal symmetric state.

\subsection{Positions of the energy minima}
\label{subsec:posenergy}
\begin{figure}
\centerline{
\epsfxsize=0.35\textwidth
\rotate[l]{\epsfbox{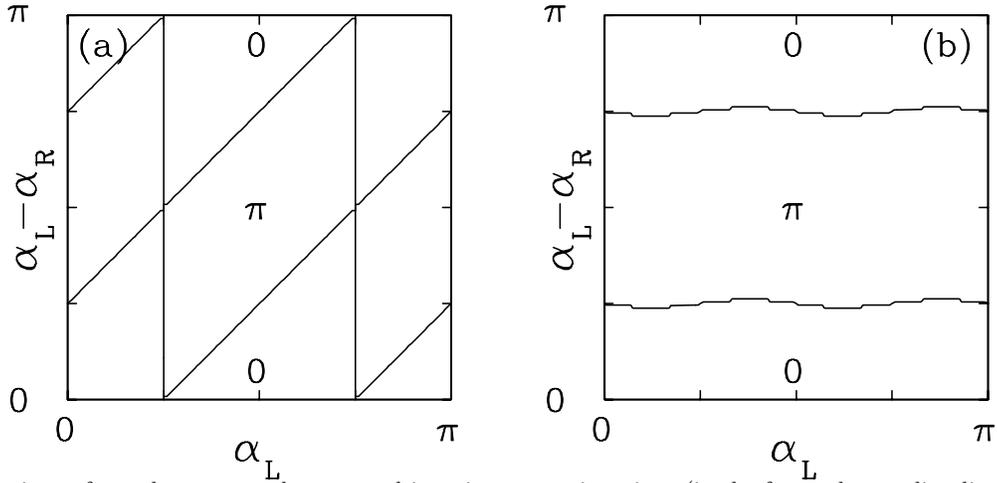}}}
\caption[]{Comparison of angular ranges where tunnel
junctions are $\pi$-junctions 
(in the forward-tunneling limit) between
$B_{1g}$, (a); equation (\ref{glcurr}),(b). }
\label{fig:gllimit}
\end{figure}

We now turn to the positions of the energy minima. 
There is an essential difference between the angular 
dependences in weak-links and tunnel junctions. 
If we ignore the change in the order parameter
near the surface, the forward (backward) tunneling limit of
eq. (\ref{tuncurrent}) is
\be
j_s(\chi)=j_0 \cos 2\alpha_L  \ \cos 2 \alpha_R  \sin \chi.
\label{SR}
\ee
This equation was first written down in  \cite{Sigrist92},
and frequent use of it has been made in the literature
(e.g. \cite{Tsuei94,Chaudhari94}).  It can be rewritten as
\begin{equation} 
\label{srcurr}
j_s(\chi)=j_0 {1\over 2}(\cos 2\theta +\cos 2(2\alpha_L-\theta))
 \sin \chi.
\end{equation}
with $\theta=\alpha_L-\alpha_R$.
Staring at the above (equivalent) equations, one observes that
the current-phase relation is sinusoidal which leads
to the energy minima of the junction always sitting at the values
$0$ or $\pi$ of the phase difference. The rest in the
equations can at most change the sign of the current
which cannot alter the conclusion that we are always dealing
with a normal or a $\pi$-junction.  As above the phase of
the minimum of energy is referred to as	$\chi_{min}$. Fig.
\ref{fig:gllimit}a displays the areas in which $\chi_{min}$
takes each of its possible values in terms of the crystal angles.
With pure $B_{1g}$ this plot is correct at all temperatures
also for the self-consistent solution.  It is
obvious, as well, that the critical current can only change sign
where it vanishes, i.e., when
$\alpha_L$ or $\alpha_R$ equals $\pi/4$, $3 \pi / 4$ etc.

Now take a look at the pinhole, first 
at the Ginzburg-Landau limit. At
that limit  and with a constant d$_{x^2-y^2}$ order parameter up to
the junction,  equation (\ref{gencurrent}) takes the form
\begin{equation} 
\label{glcurr}
j_s(\chi)=j_0  (\cos 2\theta-{1\over15} \cos 2(2\alpha_L-\theta))
\sin \chi
\end{equation}
The argument above that leads to $\chi_{min}$ always equal to
$0$ or $\pi$ applies here as well. Because of the different
coefficients in front of the second cosine terms, 
Eqs.(\ref{srcurr}) and
(\ref{glcurr}) deliver very different zones of normal or $\pi$
behaviors, as highlighted in Fig. \ref{fig:gllimit}. 
Equation  (\ref{srcurr}) assigns the junction orientation an
overestimated weight in positioning the $\pi$ behavior compared 
to 
equation (\ref{glcurr}) which emphasizes the 
misorientation of the two superconductors. This difference 
can be ascribed to the pinhole's allowing a larger incident angle 
for quasiparticles to be transported 
across the junction.  Although illustrated with the 
non self-consistent order-parameter junction
we shall make it clear later that the result is much more general.

\begin{figure}
\centerline{ 
	\epsfysize=0.4\textwidth \rotate[r]{
	\epsfbox{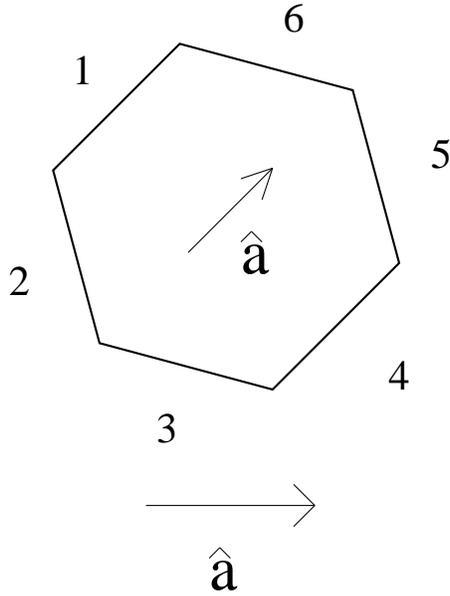} 
                         }
}
\caption[]{Hexagonal inclusion of \cite{Chaudhari94}.
The inside has the $\hat a$
axis rotated through $\pi / 4$ with respect to the outside.
$\hat a_{in}$ is roughly parallel 
to the sides $1$ and $4$ in \cite{Chaudhari94}}
\label{fig:hexa}
\end{figure}

\begin{figure}
\centerline{
\epsfxsize=0.35\textwidth
\rotate[l]{\epsfbox{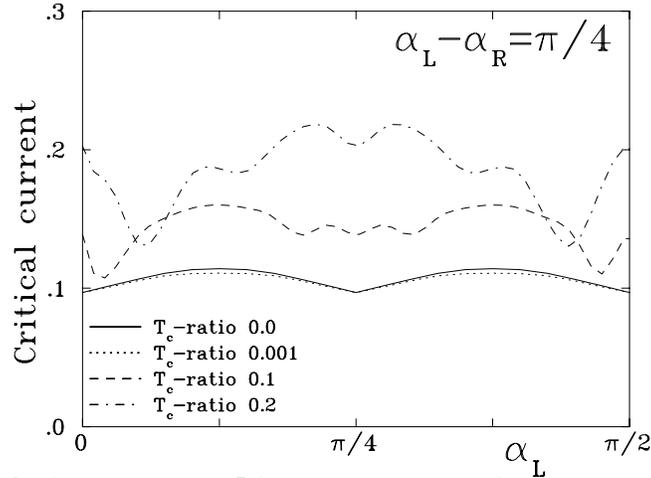}}}
\caption[]{Critical currents at a fixed misorientation ${\pi
\over 4}$ but varying junction direction $\alpha_L$ of a weak link
at $T=0.1 T_c$. The pattern has the period  ${\pi\over 2}$.}
\label{fig:ccjmax}
\end{figure}

Chaudhari and Lin \cite{Chaudhari94} measured the critical current from
a hexagon of $d_{x^2-y^2}$-superconductor to another
crystal of the same superconductor in which the hexagon was
inbedded. The misorientation between the two is 
$\alpha_L-\alpha_R ={\pi\over 4}$. 
The interfaces were the edges of
the films which the authors interpreted as tunnel junctions.
At the different boundaries of the hexagon the angles
$\alpha_L$ were equal to ${\pi\over 4}$,$-{\pi\over 12}$ etc.
(see Fig. \ref{fig:hexa})
Eq. (\ref{srcurr}) gives the critical currents across the
edges of the hexagon as $j_c=0,\pm 0.433 j_0$.
This is in contradiction with the
measured currents, which display a modulation around an average
current. The result of the experiment has been cited as evidence
against the d$_{x^2-y^2}$-pairing state for the cuprates (see for
instance \cite{Sigrist95}). Taking the crystal contacts for weak
links, the results of Chaudhari and Lin do not  rule out the
d$_{x^2-y^2}$-pairing state. In Fig. \ref{fig:ccjmax} the
critical current at $T=0.1 T_c$ for various $T_c$-ratios are
plotted. The critical
current is taken at the constant misorientation ${\pi \over
4}$. Picking the simplest case with no subdominant 
components the
critical current oscillates around an average, just as in the
experiment. The amplitude of the modulation is admittedly 
smaller than experimentally 
reported. Obviously the pinhole is a very idealized
model for a weak link, and we cannot expect 
every small detail getting accounted for.

\begin{figure}
\centerline{
\epsfxsize=0.7\textwidth
\rotate[l]{
\epsfbox{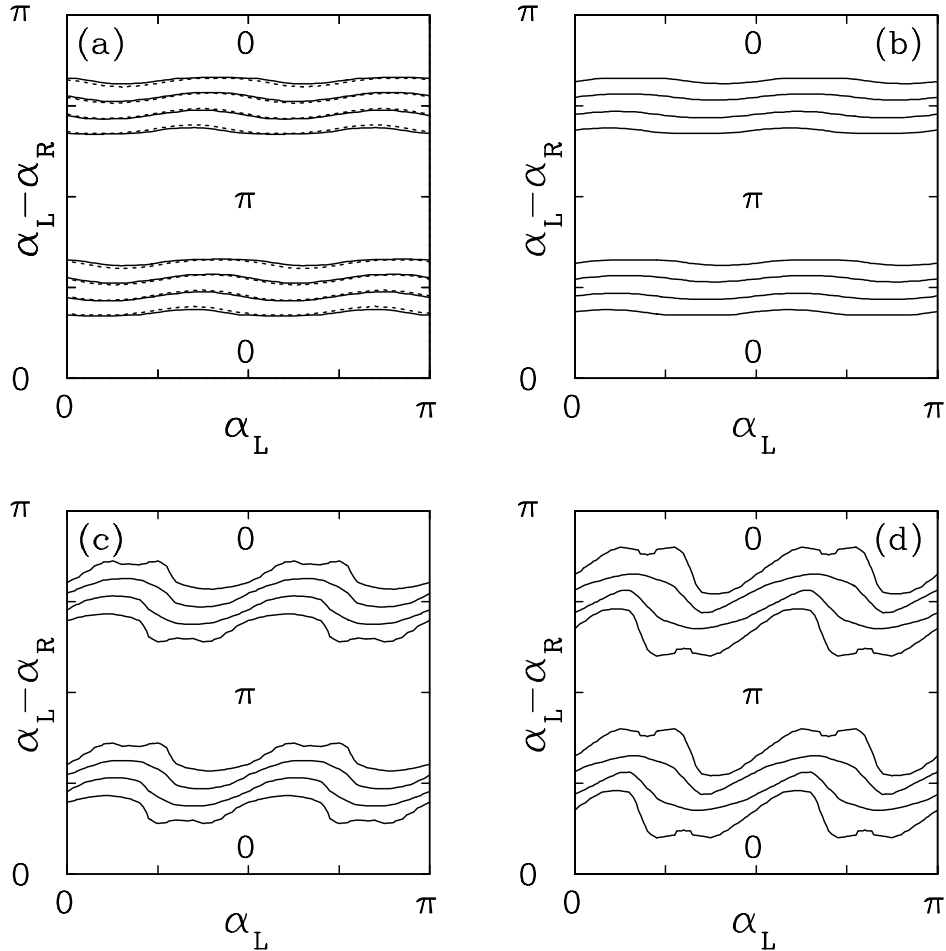}}}
\caption[]{Contour map of energy minima at
$T=0.1T_c$ in a weak link. 
Finite slopes are where the junction is neither
a 0-junction nor a $\pi$-junction. Full lines, 
self-consistently calculated results, dashed line in
(a), constant-order-parameter-up-to-the-wall-model.
pure $B_{1g}$,(a); $T_c$-ratio $10^{-3}$, (b); $0.1$,(c); 
$0.2$(d) .
With the larger $T_c$-ratios, the
junction orientation acquires an increased importance 
as the amplitude of
the $B_{2g}$-component depends on the interface orientation.}
\label{fig:wleminima}
\end{figure}

\begin{figure}
\centerline{
\epsfxsize=0.7\textwidth
\rotate[l]{
\epsfbox{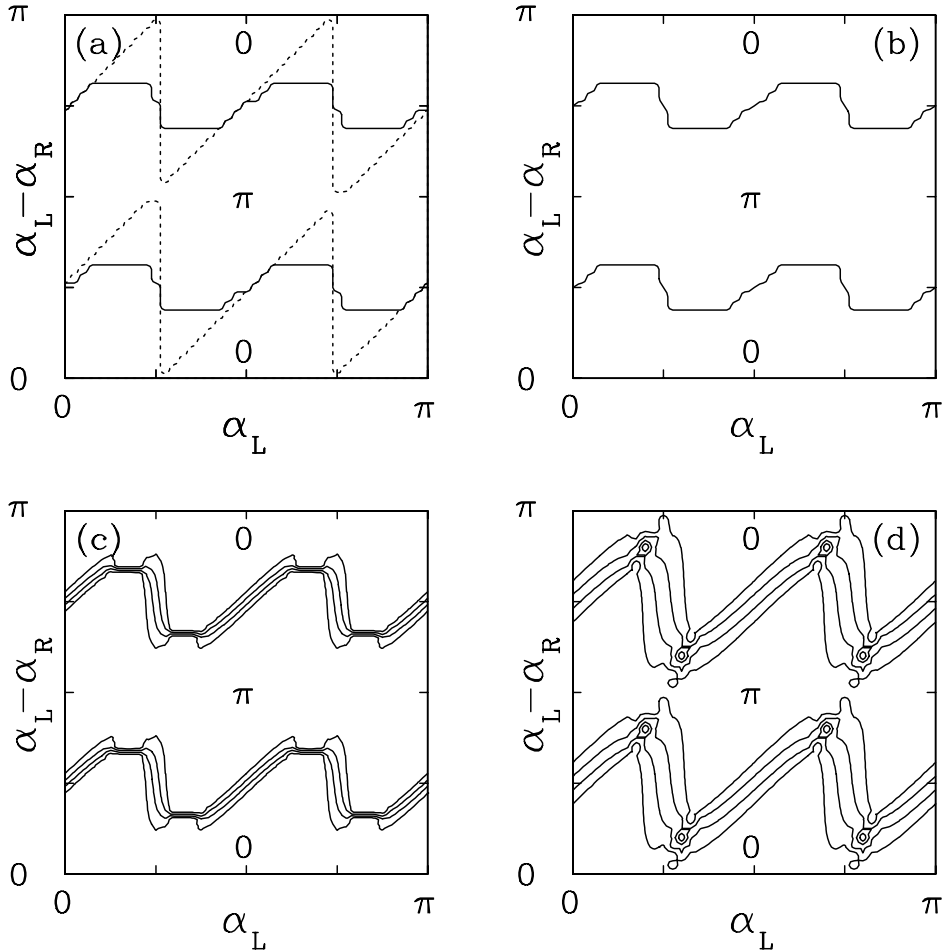}}}
\caption[]{Contour map of the positions of the energy minima at
$T=0.1T_c$ for a tunnel junction. Full lines indicate 
self-consistent results, dashed line in (a) 
constant-order-parameter model. (a)
pure $B_{1g}$ superconductors 
(b) $T_c$-ratio = $10^{-3}$,  (c) $0.1$, (d) $0.2$ 
(The difference between the
full and dashed lines in (a) arises due to the existence of zero energy
bound states near the surfaces in those orientations.
see \cite{Tanaka,Barash96}) }
\label{fig:tleminima}
\end{figure}

Away from the Ginzburg-Landau region,
the positions of the energy minima were studied numerically. 
$T=0.1 T_c (B_{1g})$ was chosen. In Figure \ref{fig:wleminima}
the position of the energy minima $\chi_{m}$ at a given crystal
misorientation and a given junction orientation can be seen
for the weak link.
The corresponding graphs for the tunnel junction can
be found in
Fig. \ref{fig:tleminima}. For comparison, the junction with a
constant order parameter up to the wall without a subdominant
component is included. Only one of the two phase
differences $(\chi^{(1)}_m,\chi^{(2)}_m)$  related through
$\chi^{(2)}_{m}=2\pi-\chi^{(1)}_{m}$ is displayed.
   
The constant-order-parameter weak-link junction has zones where
$\chi_m$ continuously evolves from zero to $\pi$. A self-consistently
determined order parameter leaves this picture unaltered, i.e. it does
not lead to noticeable changes in the positions of the weak-link
energy minima  at zero $T_c$-ratio. This is in contrast to the tunnel
junction  (not in the TRSB-state) where the changes continue being
abrupt and the constant-order-parameter model and the self-consistent
junction deliver very different pictures (see
Fig. \ref{fig:tleminima} a) . 
In the weak link,  $\chi_{min}$ is a strong
function of the misorientation $\alpha_L-\alpha_R$ but depends
only weakly
on $\alpha_L$ (or $\alpha_R$) separately. 
The tunnel-junction $\chi_m$  varies as a function of 
both the misorientation and the
junction direction. This is true as long as the admixture of the
$B_{2g}$-component is kept small (in Fig. \ref{fig:wleminima}.b
$T_c$-ratio is $10^{-3}$). Cranking up the $T_c$-ratio (see
Figs. \ref{fig:wleminima}.c-d.) adds structure to the
boundary areas between normal and $\pi$ behaviors of both 
junction types. Most structure is found when the
superconducting state close to the junction breaks time-reversal
symmetry with the mixture $d_{x^2+y^2}+i d_{xy}$ as the order
parameter. The TRSB also smoothes out the jump between
the zero junction and
the $\pi$-junction areas in the tunneling limit. 

Higher temperatures only make narrower
the zones of continuous change in $\chi_m$ from zero to
$\pi$. Above 0.5 $T_c$ this area has essentially vanished and the
Ginzburg-Landau formula (\ref{glcurr}) is an amazingly
accurate description of the weak link. 

%
%
\section{Experimental implications}


\subsection{ Small spontaneous fluxes at crystal interfaces}

An important implication of the present results bear on
the interpretation of the experiment of \cite{Kirtley95}. 
In this experiment Kirtley et al manufactured
triangular or hexagonal inclusions where the 
region inside was $45 \deg$ misoriented with respect to
the outside (similar to that of \cite{Chaudhari94}, 
see Fig. \ref{fig:hexa} and Section 5.2). Fluxes spontaneously 
generated at the corners were measured with a SQUID. The fluxes 
found were usually small fractions of the flux quantum $\Phi_o$, 
neither integer nor half-integer multiples. Moreover, the crystal 
orientations themselves (with respect to the normal of the junction 
plane) seem to have no obvious correlation with the 
fluxes observed (Kirtley, private communication).

When the dimensions of the inclusion for the measurement of flux 
are much larger than the Josephson penetration length $\lambda_J$,
the flux generated at a corner
should be ${\Phi_o \over 2 \pi} ( \chi_{min}^a - \chi_{min}^b)$
where $\chi_{min}^{a,b}$ are the position of the energy minima
of the two interfaces meeting at that corner
(see, e.g.\cite{Millis94}). 
If one assumes junctions, such as the tunnel junction,
which display 
$\chi_{min}$ $=0$ or $\pi$ as long as the order parameter
does not break the time reversal symmetry, then 
the experimental results
force one to postulate TRSB at least at the grain
boundaries \cite{Sigrist95b}.  Even with
TRSB, the observation that the crystal orientation
themselves seem to have no significant effect on the 
fluxes observed would still
find no explanation. If the bulk order parameter is pure
$B_{1g}$ then a TRSB order parameter near the interface
can only occur for $\alpha_L$ or $\alpha_R$ near $\pi / 4$,
whereas $\alpha_R - \alpha_L$ is not directly relevant
(c.f. Fig. \ref{fig:gllimit} a.)

If the interfaces are weak-links, however, everything falls
in place.  Since $\chi_{min}$ is neither $0$ nor $\pi $ in
general even without TRSB (see Fig. \ref{fig:wleminima}).
Our unit-transmission limit seems to be somewhat too extreme, 
however, in that
$\chi_{min}$  depends on $\alpha_L-\alpha_R$ 
but is almost independent
on the individual $\alpha_L$ or $\alpha_R$. For pure $B_{1g}$, the
theoretical fluxes are too small. This holds true also with
a substantial $B_{2g}$ component induced near the grain boundaries
as long as we are not in a TRSB state (see Fig. \ref{fig:wleminima}).
In reality the grain boundary is more likely to have 
a smaller than unity transmission coefficient, so 
that $\chi_{min}$ would vary stronger with the
$\alpha$'s (i.e, with pure $B_{1g}$, the portrait of $\chi_{min}$ 
would have to evolve from Fig. \ref{fig:wleminima} a to 
Fig. \ref{fig:gllimit} a when the transmission 
decreases).  \footnote{Details of the calculations
with a finite transmission coefficient
will be reported separately}.

The condition $\alpha_L - \alpha_R$
approximately equal to $\pi/4$ is crucial.  For 
example other experiments
 involving misoriented
crystals, \cite{Tsuei94,Tsuei96,Kirtley96}
(as well as the "frustrated" samples of \cite{Miller95})
all have misorientations close to (but sometimes not exactly) $\pi/6$
and one of the $\alpha$'s is approximately zero.
(or those related to these by symmetry)
Our pinhole results for $\chi_{min}$ 
with the transmission coefficient equal
to unity,
differ only very slightly 
from the corresponding tunnel-junction values.
( For the junctions with $\alpha_L=0, \alpha_R = \pm \pi / 6$, 
$\chi_{min} \approx 0.17 \pi$
 \footnote{ this value is for
at $T/T_c = 0.1$, slightly higher than 
that appropriate for some of the experiments
( $T \approx 4.2 K$, $ T_c \approx 90K$). }
rather than  $\chi_m = 0$ for tunnel junctions without broken 
time reversal symmetry.
).
The deviation for these new energy minima will be even smaller
with a smaller transmission coefficient and/or
more directional tunneling. 
Thus our picture for the small
spontaneously fluxes of \cite{Kirtley95} is not inconsistent
with all these experiments.
It would be interesting experimentally to
try to observe the fluxes, if any, that nucleate
at the corners of the inclusions with misorientation of
$\pi/6$ with judicious choice of $\alpha$'s. 
 Obviously the results expected from the energy minima
based on weak-links (Fig \ref{fig:wleminima}) will be
very different from those based on eq(\ref{SR})
(Fig \ref{fig:gllimit} a) 
as well as those from tunnel junctions with
TRSB near the interfaces (Fig \ref{fig:tleminima} c,d).

\subsection{The transmission coefficient of a junction}
$J_c$ across grain boundaries have been measured.
They range from $ \sim 10^{3} A /cm^2$ to $\sim 10 ^{5} A/cm^2 $
for $ T << T_c$ ( e.g. \cite{Miller95}, \cite{Chaudhari94} 
\cite{Tsuei94}, \cite{Tsuei96}, \cite{Kirtley96}). 
A transmission coefficient can be inferred
from these values.  The critical
current for a weak-link is dictated by the formation of
phase slip centers and is not equal to the thermodynamic
current directly related to the transmission coefficient.
Recently \cite{Tsuei96} and 
\cite{Kirtley96} have reported  direct 
measurements of the Josephson penetration depth $\lambda_J$
via the observation of fluxes along the grain boundaries
and the corner of the tricrystals (with a misorientation of
$\sim 30 \deg$).  We shall try to estimate
the transmission probability $\vert T \vert^2$ from their 
$\lambda_J$. With $\lambda_J^{-2} = { 8 \pi J_c \over \hbar c^2} w$
where $w$ is the thickness of the magnetic field penetration
layer along the barrier, and with the rough formula
$J_c \sim \pi \Delta N_f v_f e \vert T \vert^2 $ for
the critical current density,  as well as $\lambda_L^{-2} 
= 4 \pi N_f v_f^2 e^2 / c$ for the London penetration length,
we arrive at ${ \lambda_L^2 \over \lambda_J^2} \sim { w \over \xi_0} 
\vert T \vert^2 $ where $\xi_0 = { \hbar v_f \over \pi \Delta}$
is the zero temperature coherence length.  For $YBCO$, 
if we take $\lambda_L \approx 1400 {\rm \AA}$ , 
$\xi_0 \approx 14 {\rm \AA}$,
with $\lambda_J \sim \mu m $ and $ w \approx 2 \lambda_L$, 
we get $\vert T \vert^2 \sim 10^{-4}$. Estimates 
for $TlBaCuO$ are similar.  At first sight then, $\vert
T\vert^2$ seems small enough in the $30 \deg$ misoriented films 
for one to assume the tunneling limit. If we accept the same order of
magnitude for $\vert T \vert ^2$ in \cite{Kirtley95}, then our
scenario for the new $\chi_m$ seem not very plausible.  One needs to
keep in mind, however, that the estimate is an average transmission 
coefficient.  Electrical conduction may proceed through
"hot spots" of much higher transparency.
If we take the extreme limit of the  pinholes with unit transmission,
then the fraction of transparent area is $\approx 10^{-4}$. 
It is possible that the non-uniformity of
the critical current densities cannot be seen in the 
scanning SQUID measurements if the distances between the
pinholes are $\ll \lambda_J$. If $d$ is the average pinhole-pinhole
distance and $(a/d)^2 \sim 10^{-4}$, there remains reasonable
margin for the parameters $a$ and $d$ even with the restriction
$ a << \xi_0 << d << \lambda_J$ for our calculation to apply.
In \cite{Kirtley95}, the authors themselves 
emphasized that the fluxes would be
much more localized than expected if $\lambda_J$ were
$\sim \mu m$, which may suggest that the boundaries are
very non-uniform. 

\section{Conclusion}

In conclusion we have investigated the difference between
high and low transmission junction in d-wave superconductors,
using pinhole as an extreme example in the high transmission
limit.  The effects of subdominant order parameter were
also considered.

\vskip 0.5 cm

\noindent{\bf Acknowledgement}

This research was supported in part by the NSF through the
Northwestern 
University Materials Research Center, grant no. DMR 91-20521 (SY),
the Science and Technology Center for Superconductivity , 
grant no.  DMR 91-20000 (MF and SY), 
and Academy of Finland under contract No. 1081066 (MF) and 
on a sabbatical leave (JK).
MF also acknowledges partial support from SF{\AA}AF,
 {\AA}bo Akademi and Magnus Ehrnrooths Stiftelse.
Two of us (JK and SY) would like to thank Aspen Center for Physics,
where this collaboration was initiated.

%
%

%
%
\end{document}